\documentclass[twocolumn,showpacs,preprintnumbers,amsmath,amssymb]{revtex4}

\usepackage{graphicx}
\usepackage{dcolumn}
\usepackage{bm}

\def\LSCO{La$_{2-x}$Sr$_x$CuO$_4$}

\begin{document}

\preprint{}

\title{What the resonance peak cannot do}
\author{Hae-Young Kee$^{1,2}$, Steven A. Kivelson$^{1}$, and G.
Aeppli$^{3}$}
\affiliation{$^1$Department of Physics, University of California Los
Angeles,  
Los Angeles, California 90095--1547\\
$^2$ Department of Physics, University of Toronto, Toronto, Ontario M5S
1A7, Canada\\
$^3$NEC Research Institute, 4 Independence Way, Princeton, NJ 08540}
\date{\today}

\begin{abstract}

In certain cuprates, a spin 1 resonance mode is prominent
in the magnetic structure measured by neutron scattering.  It has
been proposed that this mode is responsible for significant features
seen in other spectroscopies, such as photoemission and optical
absorption, which are sensitive to the charge dynamics, and even 
that this mode is the boson responsibile for ``mediating'' the
superconducting pairing.  We show that its small (measured) intensity and
weak coupling to electron-hole pairs (as deduced from the measured
lifetime) disqualifies the resonant mode from either proposed role.
\end{abstract}


\maketitle

\paragraph{Introduction}

One of the most striking features
of the high temperature superconducting cuprates is the sharp resonance
peak observed  in inelastic neutron scattering
measurements.\cite{resonance,tranquada,mook,keimer,bscco,fong} This
phenomenon has a clear and
intimate relation to the superconductivity that occurs in these
materials -
the resonance grows in intensity and narrows at
temperatures less than the superconducting $T_c$, and its intensity
is suppressed by perpendicular magnetic fields in a way
that correlates directly with the suppression of
superconductivity\cite{dai}.  

Since its discovery, there have been
many interesting theoretical proposals concerning the origin and
implications of the resonance.
One class of proposals identifies the resonance peak as a
signature of superconductivity, relating its intensity to
the condensation energy\cite{Zhang98} or condensate
fraction\cite{Chakravarty00,Zhang98} of the superconducting state. Other
proposals
focus on the {\it effect} that scattering of quasi-particles from the
resonance has on various other experimentally accessible properties of
the cuprates, especially those that show dramatic changes as the
temperature is lowered from above to below $T_c$. 
For instance, this idea has been invoked to explain the
``peak-dip-hump''
structure\cite{Norman00,abanov} and the ``kink''\cite{Valla} in the
quasi-particle dispersion measured by angle resolved photoemission
spectroscopy (ARPES), and the pseudogap structure seen in optical
conductivity\cite{Carbotte01}. Finally, there are proposals which
consider the resonance mode to be the boson which ``mediates'' an
effective attraction between electrons which is responsible for the
high-temperature superconducting pairing
\cite{carbotte99,orenstein,abanov}.  

In this paper, we address what the
resonance mode can and cannot do.  In particular, we will show that the
first set of ideas
requires that the integrated intensity associated with the resonance,
$I_0$, be small in units of the total integrated spin structure
factor, while the latter require $I_0\sim1$.  While
{\it apparently} contradictory numbers exist in the experimental
literature, a careful analysis\cite{hayden88,dai2,aeppli} shows that 
values of $I_0$ in the few percent range can be deduced from {\it all}
the
absolute intensity measurements.  (See Table I.)  Thus, the resonance
peak may be a
unique {\it probe} of the superconducting
condensate\cite{Chakravarty00}, and might account for the condensation
energy\cite{Zhang98,dai2}  but it cannot  {\it cause} any
significant 
structure in ARPES and optical conductivity. 

\paragraph{Possible Relation to the Condensation Energy}
The concept of condensation energy is not well-defined when fluctuation
effects are important\cite{sudip2}.
In the absence of better estimates, we adopt the mean-field expression
for the condensation energy:
\begin{equation}
U= \frac{1}{2} \rho \Delta^2 \sim 2 \rho T_c^2.
\end{equation}
If the density of states is proportional to $1/(2 J)$ where $J$
is the exchange interaction in the $t-J$ model invoked in Ref.
\cite{Zhang98},
the condensation energy can be expressed as
\begin{equation}
U \sim \frac{T_c^2}{J}= J \left( \frac{T_c}{J} \right)^2.
\end{equation}
In this context, White and Scalapino\cite{scalapino} have pointed out
that there is an exact relation between the nearest-neighbor exchange
contribution to the internal energy and the magnetic structure factor
$S(\vec k,\omega)$ which in spatial dimension $d=2$ is
\begin{equation}
U_{mag}= J \int {d\omega d\vec k\over (2\pi)^{d+1}}
[2-\cos(k_x)-\cos(k_y)]S(\vec
k,\omega).
\end{equation} 
Therefore, it is clear that {\bf if} the condensation energy comes
principally from
this term, and {\bf if} it is due to the transfer of spectral weight
from a broad (in
$\vec k$) background into the resonant  peak, {\bf then} its intensity
must be very  {\it
small}, $I_0 \sim (T_c/J)^2$.
Also in accord with these ideas, Dai et al \cite{dai2}
have shown that, in absolute units, the specific heat is roughly equal to
the temperature derivative of the resonance intensity.

\paragraph{Scattering from a collective mode}
A collective mode with a strongly temperature-dependent intensity is
unusual
, so it is natural to attribute other strongly temperature
dependent spectral features to the coupling between quasi-particles and
the
collective mode.  However, unless the mode has large weight, it is as if
it were
hardly there at all, however prominent it may appear in a scattering
experiment. 
Consider, for example, the electron-phonon coupling in a weakly
correlated metal
with a very large number, $N$, of atoms per unit cell.  The scattering
of electrons
from any one optical mode will not, generally, have
significant
effect on the electron dynamics - its effects will be reduced by a
factor of
$1/N$.  

Several prominent features of the ARPES spectra of the high
temperature superconductors have been attributed to scattering of
electrons from
the resonant mode, in particular the
pronounced peak-dip-hump structure in the ``antinodal region'' of the
Brilloin zone
(near
$\vec k=<\pi,0>$)  and the kink in the electron dispersion seen
especially in the ``nodal direction'' ($<0,0>$ to $<\pi,\pi>$).  These
are order 1
effects, and so require a {\it large} intensity of the resonance peak
unless the coupling to
quasiparticles is extremely large\cite{Norman00}. 
On the same grounds, the pseudogap features in the optical conductivity
require a large scattering from the resonant peak. 

Therefore, it is clear that the resonance peak cannot be
responsible for both the condensation energy {\it and} the pronounced
structures in the scattering rates.


\paragraph{The spectral intensity of resonance peak is small!}
The resonance peak is the most prominent feature seen in inelastic
neutron scattering in the poster-child of high temperature
superconductors,
YBCO\cite{resonance,tranquada,mook, keimer,dai2}. Even though it turns
on at a temperature which increases with decreasing $T_c$ and thus
tracks the celebrated pseudogap phenomenon, its most rapid evolution in
intensity, lifetime, and width in k-space occurs near $T_c$. 
In addition, its frequency scales with $T_c$ and its
field-dependence\cite{dai}, both in magnitude and anisotropy, is linked
to the upper critical field H$_{c2}$. These experiments establish
a strong connection between the resonance and both the spin and orbital
aspects of  superconductivity.
However, the resonance is so prominent only because its spectral weight is
concentrated in a narrow range of frequency and $\vec k$.

Absolute intensity measurements reveal that the intensity, when properly
integrated over frequency and the Brillouin zone, is always
rather small\cite{aeppli}.  (See Table I.)  Simple considerations of the
chemistry and physics of the copper-oxide planes leads to the
conclusion that each planar copper is in a
$d^9$ configuration with its orbital angular
momentum quenched by the crystal field, leaving only a S=1/2 degree of
freedom.  We therefore expect the total integrated spectral weight per
planar copper to be
$\hbar^2S(S+1)=\hbar^2(3/4)$.  In YBCO, the
measured spectral
weight in the resonance is\cite{aeppli} of the order of 1\% - 2\% of
this.  In BSCCO, because
the resonant peak appears broader\cite{bscco} in $\vec k$, its
integrated strength might be somewhat larger,
but still at the 5\% level.    These ratios are not subject to
significant uncertainties.  For example, the same experimental
methods show that in the undoped,
antiferromagnetic ``parent'' materials, the measured spectral
weights\cite{hayden,coldea}
are in quantitative agreement with the results of spin-wave theory for a
S=1/2 system. 
Furthermore, the relatively low doped hole densities, even at ``optimal''
doping, implies that the total magnetic spectral weight (below the charge
transfer gap) cannot differ greatly from S(S+1).

\begin{table}
\begin{center}
\caption{Integrated spectral weight in the resonant peak well below
$T_c$ in units such that a
spin 1/2 per planar Cu atom would have integrated weight equal to 1.}
\begin{tabular}{|c||c|c|c|}
\hline
{\large Material} & {$T_c$ (in K) } &
{$I_0$} & {Reference} \\
\hline
\hline
{YBa$_2$Cu$_3$O$_{6.5}$} & {52} &
{0.017} & {\cite{fong}}\\
\hline
{YBa$_2$Cu$_3$O$_{6.6}$} & {62.7} &
{0.01$\pm 0.007$} & {\cite{dai2,aeppli}}\\
\hline
{YBa$_2$Cu$_3$O$_{6.7}$} & {67} &
{0.014} & {\cite{fong} }\\
\hline
{YBa$_2$Cu$_3$O$_{6.85}$} & {87} &
{0.017} & {\cite{fong}} \\
\hline
{YBa$_2$Cu$_3$O$_{6.99}$} & {93} &
{0.011} & {\cite{fong} }\\
\hline
{Bi$_2$Sr$_2$CaCu$_2$O$_{8+\delta}$} & {91} &
{0.057} & {\cite{bscco}}  \\
\hline   
\end{tabular}
\end{center}
\end{table}


\paragraph{A little mathematics}
Let us now carry out the simplest calculation to illustrate
our argument.
The Hamiltonian  which represents the coupling between
the conduction electrons and the spin mode is as follows.
\begin{equation}
H= H_0 + g {\bf S} \cdot \psi^{\dagger} {\vec \sigma} \psi
+ H_s,
\end{equation}
where $H_0$ and $H_s$ are the bare Hamiltonians for the conduction
electrons and spins respectively.

We approximate the 
imaginary part of the zero temperature spin susceptibility, measured via
neutrons, as
\begin{eqnarray}
&&{\rm Im} \chi({\bf q},\omega)=(\pi/3)g_L^2\mu_B^2S(S+1) \\
&&{\times}\left\{I_0 (2\pi)[\delta(\omega-\Omega)-\delta(\omega+\Omega)] 
f({\bf q})  + \frac{(1- I_0){\rm sign}(\omega)}{\Lambda_{\omega} \Lambda_q^d}
\right\},\nonumber
\end{eqnarray}
where $\Omega$ is the resonance frequency, $g_L$ is the Lande g-factor, and
 $\Lambda_{\omega}$ and
$\Lambda_q$ are, respectively, the frequency and momentum cut-offs.
The structure factor of the resonant mode, $f({\bf q})$,
is known to be  peaked at $\vec q$ in the neighborhood of
${\bf Q}= (\pi,\pi)$;
for simplicity we will take
$f({\bf q})= (2\pi)^d \delta({\bf q}-{\bf Q})$, although the results are easily
generalized to the case in which $f$ is a Lorentzian or Gaussian.  
We will also 
take $d=2$, although, of course, the real cuprate superconductors are
anisotropic  three dimensional systems.

\subparagraph{ARPES}
The leading perturbative contribution to the self energy from the
resonance
peak is written as
\begin{equation}
\Sigma({\bf k},\omega) =  I_0 g^2 \int d^2 q
\left( \frac{1}{\omega-\Omega-\xi_{{\bf k}-{\bf q}}+ i\eta} \right)
\delta({\bf q}-{\bf Q}),
\end{equation}
where $\xi_{{\bf k}}$ is the quasi-particle dispersion.
Therefore, the single particle spectrum has two poles located at
\begin{eqnarray}
\omega_1 &=& \xi_{\bf k} -\frac{I_0 g^2}{\Omega} + \ldots
\nonumber\\
\omega_2 &=& \Omega+\xi_{{\bf k}-{\bf Q}} +\frac{I_0 g^2}{\Omega}
+\ldots,
\end{eqnarray}
where $\ldots$ refers to terms of order ${\cal O}(\frac{I_0^2
g^4}{\Omega^3})$.
The weight of each pole is
\begin{eqnarray}
Z_{\omega_1} &=& 1-\frac{I_0 g^2}{\Omega^2}+\ldots
\nonumber\\
Z_{\omega_2} &=& \frac{I_0 g^2}{\Omega^2}+\ldots
\end{eqnarray}
To the same order, the scattering from the remaining (non-resonant)
spin
fluctuations produces an additive
contribution to
$\Sigma$ proportional to
$g^2(1-I_0)$ which is of the  marginal Fermi liquid form, discussed
elsewhere\cite{mfl}.

\subparagraph{Optical conductivity}

While a perturbative expression for the conductivity, $\sigma(\omega)$,
itself is impossible, due to its singular behavior at small $\omega$, it
is 
straightforward to obtain a perturbative expression for the so-called
frequency-
dependent scattering rate\cite{taustar}, defined in terms of the real
and imaginary parts,
$\sigma'$ and $\sigma''$,
as
\begin{equation}
1/\tau^*(\omega)\equiv \omega \sigma'/\sigma''=
1/\tau^*_0(\omega)+1/\tau^*_1(\omega) .
\end{equation}
To lowest order,
the contribution to the $T=0$ scattering rate from the resonance mode
is
\begin{equation}
1/\tau^*_{0}(\omega) = \frac{m \omega^2} {n e^2} g^2 I_0F(\omega),
\end{equation}
where $4\pi e^2n/m$ is the Drude weight and
\begin{equation}
F(\omega) =  \frac{\pi^2 e^2 }{\omega^3 m^2 v_F v_{\Delta}}
(\omega-\Omega) \theta(\omega-\Omega).
\end{equation}
To obtain the explicit expression for $F(\omega)$, we have used the
dispersion
relation for the nodal quasiparticle with two different velocities,
where $v_F$ and $v_{\Delta}$ are, respectively, the velocities
perpendicular 
and parallel to the Fermi surface. 
A different assumed dispersion relation would not change the overall
conclusion of this paper, although it would change the detailed structure
of
$F(\omega)$.
The contribution of the constant part of the spin susceptibility is,
unsurprisingly,
as in a marginal Fermi liquid, linear in the frequency
\begin{equation}
1/\tau_{1}^*(\omega) \propto g^2(1-I_0) |\omega|.
\end{equation}
Since Matthiessen's rule holds to this order, these scattering rates
should simply be added (and so should the scattering due to any other
process).

Our analytic results are consistent with the more complicated results
obtained
for more detailed and realistic models
previously.\cite{Norman00,Carbotte01}
However, the present results highlight the fact that all effects of the
resonance
mode are proportional to $I_0$, and so are negligible if $I_0$ is small.

\paragraph{What about the coupling constant?}
The effects of the resonant mode are not just proportional to $I_0$,
but depend on the coupling strength $g$.  Could we imagine obtaining a
large effect with a small $I_0$ but a large $g$?  Of course, a
 large $g$ is incompatible with any sort of perturbative
treatment, so such an approach probably does not make sense.
However, it also turns out that one can obtain a reasonable
estimate of the coupling constant from the experimentally measured
frequency width
(lifetime) of the resonance peak. The resonance mode in YBCO in
the superconducting state is very sharp, with an intrinsic line width
$(\Gamma)$ of about $ 2 meV$. In optimally doped material, the mode is
unobservable above $T_c$, but in underdoped material it persists to
higher
temperatures.  Here, the line broadens\cite{dai2} so that $\Gamma\approx
10 meV$.
Such a broadening is expected whenever
the resonance mode can decay into
electron-hole pairs. This decay channel is somewhat suppressed well
below
$T_c$ due to the limited phase space available for such electron-hole
pairs.   

An analogous problem was solved long ago, where instead of the resonant
mode, 
the crystal-field excitations in metallic rare-earth systems were
investigated.  Simply adopting the expressions obtained there for the
damping 
of an electronic collective mode due to electron-hole excitations, one
obtains
\cite{Fulde77,Feile81}
\begin{equation}
\Gamma= 4 \pi [g N(0)]^2 \Omega,
\end{equation}
where $N(0)$ is the density of states at the Fermi energy, or more
precisely the
density of particle-hole states with momentum $\vec Q$ and energy
$\Omega$. In order to invert
this equation to obtain an estimate of $g$, we can use  the measured
values of $\Gamma$ and $\Omega$, but we need an estimate of $N(0)$.  In
the normal state, this
can be done in several ways.  Firstly, on the basis of the theoretical
expectation that the
bandwidth of the electrons is renormalized down to something of order
$J$, or for that matter
from the measured ARPES spectra, it is possible to obtain a rough
dimensional estimate of
the normal state density of states,
$N(0)= 1/W \sim 1/(100 meV)$, to obtain an estimate of
$g$:
\begin{equation}
g \sim 14 meV.
\end{equation}
This is not a large coupling!
The conventionally defined dimensionless coupling constant
$\lambda= 2 I_0g^2 N(0)/\Omega$ is only
$ \lambda \sim I_0/10$.  Needless to say, such a feeble (small $I_0$)
boson  coupled so weakly (small $g$) to electron-hole pairs cannot
mediate a strong pairing interaction;
searches for the mechanism of high $T_c$ must begin elsewhere. 

One might worry that our estimate of $N(0)$ is somewhat too large, as it
does not take into 
account
any suppression of the density of states due to the pseudo-gap observed
above $T_c$
in underdoped materials
- a smaller assumed
$N(0)$ would give rise to a larger estimate of $g$.  However, a
more direct estimate
of the density of states can be obtained from the measured\cite{loram}
specific heat in the
normal state; for YBCO, $\gamma\equiv C_v/T$ approaches a normal state
value of around 2 mJ/gm-at K$^2$,
which corresponds to a density of states per copper of $N(0)=11$
eV$^{-1}$, in good agreement
with our dimensional estimate.

Finally, an independent estimate of the coupling constant can
be obtained from the measured lifetime in the superconducting state.
Here, the particle-hole
continuum is dominated by the nodal quasi-particles, whose dispersion is
presumably known.  It
is straightforward to see that the appropriate density of states with
momentum $\vec Q$ and
energy $\Omega$ computed within this model is $\omega/(v_F v_{\Delta}
k_n^2)$, 
where $\vec k_{node}=<k_n,k_n>$ is the
position of the nodal point measured from $(\pi/2,\pi/2)$.  
With this expression for the density of states, and taking
the canonical values of $v_{\Delta} \sim 1.2 \times 10^6
cm/s$\cite{Taillefer}, 
$v_F \sim 1.7 \times 10^7 cm/s$, and 
$k_n \sim 1/(8 \times 10^{-8} cm)$\cite{Valla2}, we obtain an estimate
$g \sim 5 meV$ and $\lambda \sim 0.03I_0$.  

Although it takes us a bit into the realm of speculation, it is worth
noting that the
remarkably small value of the coupling to the resonant mode is not,
altogether, unexpected.  If
we think of the resonant mode, in some loose sense, as a would-be
antiferromagnetic
magnon\cite{abanov}, then an argument due to Schrieffer\cite{schrieffer}
implies that it couples only through
gradient couplings to particle-hole pairs.  In particular, one might
expect
the average coupling to be roughly proportional to the reciprocal space
width around $\vec k=\vec Q$ occupied by the resonant peak. 
Since this width
is of order 20\% of the width of the Brillouin zone, it is reasonable to
expect the coupling to  the {\it antiferromagnetic} resonance itself to be
correspondingly reduced relative to an order 1 microscopic coupling between
electrons and spins.

\paragraph{Conclusion}
The resonance mode is important because it is the most prominent
feature of an especially simple correlation function. It is one
of the salient features of high temperature superconductivity
whose understanding will eventually
result in significant insight into the mechanism of high temperature
superconductivity.
However, to the best of our knowledge, its spectral weight
is always
small.
Therefore, the existence and character of the
resonance
mode may well be a direct {\it consequence} of the high temperature
superconductivity in the
cuprates but it cannot be the ``glue'' in any conventional pairing
theory, nor can it account for anomalies in photoemission and optical
absorption data. This conclusion is
reinforced by the observation that many\cite{shen,basov2} of the
putative
signatures of scattering from the resonant peak are observable in
{\LSCO},  where no resonant mode has been seen in neutron
scattering.

\paragraph{Acknowledgements:}
We have benefitted greatly from helpful comments from D.Basov,
S.Chakravarty, P. Johnson, Y-B Kim,
K.Moler, C. Nayak, D. J. Scalapino, J.Tranquada, A. Tremblay, S-C.
Zhang, and especially L. Pryadko. 
This work
was supported by the Canadian Institute for Advanced Research (HYK), NSF
grant \# DMR-0110329 and
DOE grant \# DE-FG03-00ER45798 at UCLA (SK and HYK).


\begin{references}
\bibitem{Valla} P. D. Johnson {\it et al}, cond-mat/0102260.  Here, in
fact, only the
{\it change} the 10\% {\it change} in the kink in the dispersion upon
entering the
superconducting state is identified with the existence of the resonant
mode.
\bibitem{taustar}  The more commonly discussed\cite{basov3} scattering
rate is defined as $1/\tau
=\omega_p^2\sigma^{\prime}/4\pi|\sigma|^2$.  If $\tau$ is frequency
independent (in the Drude
model), then $\tau=\tau^*$.  More generally, $\tau$ and $\tau^*$ behave
more or less similarly -
we have presented theoretical results for $\tau^*$ because it does not
require an independent
opinion concerning the plasma frequency, $\omega_p$, but can be
expressed directly in terms of
measured conductivities.
\end{references}
\end{document}